\RequirePackage{fix-cm}
\documentclass[smallextended,natbib]{svjour3}       

\smartqed  
\usepackage{graphicx}
\usepackage[utf8]{inputenc}
\usepackage[T1]{fontenc}
\usepackage[english]{babel}

\usepackage{xcolor}

\journalname{Foundations of Science}
\usepackage[normalem]{ulem}

\begin{document}

\title{Physics is organized around transformations connecting contextures in a polycontextural world}
\titlerunning{Physics is organized around transformations connecting contextures}

\author{Johannes Falk \and Edwin Eichler \and Katja Windt \and Marc-Thorsten Hütt}%
\authorrunning{Falk et al.}

\institute{ J. Falk \at
            Department of Life Sciences and Chemistry, Jacobs University, Bremen, Germany\\\email{j.falk@jacobs-university.de}
              \and 
E. Eichler \at
            EICHLER Consulting AG, Weggis, Switzerland       
           \and
            K. Windt \at
            Global Production Logistics, Jacobs University Bremen, Germany
            \and
            K. Windt \and E. Eichler \at
            SMS Group, Düsseldorf, Germany
            \and 
            M.-T. Hütt \at
            Department of Life Sciences and Chemistry, Jacobs University, Bremen, Germany
}

\date{Received: date / Accepted: date}
\maketitle
\begin{abstract}
The rich body of physical theories defines the foundation of our understanding of the world. Its mathematical formulation is based on classical Aristotelian (binary) logic.
In the philosophy of science the ambiguities, paradoxes, and the possibility of subjective interpretations of facts have challenged binary logic, leading, among other developments, to Gotthard Günther's theory of polycontexturality (often also termed 'transclassical logic'). 
Günther's theory explains how observers with subjective perception can become aware of their own subjectivity and provides means to describe contradicting or even paradox observations in a logically sound formalism. Here we summarize the formalism behind Günther's theory and apply it to two well-known examples from physics where different observers operate in distinct and only locally valid logical systems. Using polycontextural logic we show how the emerging awareness of these limitations of logical systems entails the design of mathematical transformations, which then become an integral part of the theory. In our view, this approach offers a novel perspective on the structure of physical theories and, at the same time, emphasizes the relevance of the theory of polycontexturality in modern sciences.

\keywords{Polycontexturality \and Gotthard Günther \and Subjectivity \and Transformations}
\end{abstract}

\section{Introduction \label{sec:intro}}
The concept of Aristotelian logic is based on an objective truth of first principles, an absolute property that is independent of any subjectivity and context~\citep{halper_one_2009,irwin_aristotles_1990}. Something that is not true is necessarily false and likewise, something that is not false is true, or to say it in Aristotle's words: ``\textit{To say of what is that it is not, or of what is not that it is, is false, while to say of what is that it is, and of what is not that it is not, is true}''~\citep{david_correspondence_2016}. Since there are no other possible values than \textit{True} and \textit{False}, this logic is called a two-valued (also bivalent or binary) logic~\citep{goble_blackwell_2001}. Providing the historical foundation of mathematical logic~\citep{adamowicz_logic_2011,lukasiewicz1938logik} as well as Boolean Algebra~\citep{font_survey_2003}, Aristotelian logic has become a cornerstone of western thought and technology. In Aristotelian logic, truth is always inter-subjective. A sentence is only true if it is true for each object covered and for all subjects using the sentence~\cite[p.9]{gunther_idee_1991}. There is hence no subjectivity for truth.; anything described by Aristotelian logic is objective and there is only one, absolute truth~\cite[p.23]{klagenfurt_technologische_2016}. While this assumption was considered virtually incontrovertible for centuries, one of the first philosophers who questioned the concept of objective truth was Immanuel Kant. In his work ``Critique of Pure Reason''\footnote{Original German Title: Critik der reinen Vernunft} he states: ``\textit{The nominal definition of truth, that it is the agreement of knowledge with its object, is [here] assumed as granted; the question asked is as to what is the general and sure criterion of the truth of any and every knowledge.}''~(cited after \citep{sen_kant_1999}). Kant furthermore writes: ``\textit{In consequence of this mere nominal definition, my cognition, to count as true, is supposed to agree with its object. Now I can compare the object with my cognition, however, only by cognizing it. Hence my cognition is supposed to confirm itself, which is far short of being sufficient for truth.}''~\cite[p.557]{kant_lectures_1992}. While accepting a correspondence theory of truth, Kant does not give an objective criterion to test whether a cognition is true or false~\cite[p.215]{cleveProblemsKant2003}. He rather states that cognition is personal, hence subjective. Nevertheless, what Kant talks about is an epistemology that still builds upon the classical understanding of Aristotelian logic~\cite[p.22]{gunther_idee_1991}.

In the German Idealism, several philosophers premised on the ideas of Kant and used his theory to discuss self-consciousness or self-awareness. Fichte introduced the self-conscious \textit{Self} as ``absolute subject'' that needs to be distinguished from the \textit{I} as an object of reflection. Some years later, Hegel concluded that subject and object are essentially the same or -- at least -- cannot be separated as the Aristotelian logic enforces. Both Hegel and Fichte noticed that the ``thinking I'' often is part of the system it makes statements about, which introduces self-references and contradicts the Aristotelian assumption of an inter-subjective truth.

The ideas and theories developed during the German Idealism are mainly based on philosophical considerations and, similar to Aristotle's logic, were mainly focused on human reasoning. A mathematical treatment of logic was initiated by Boole and de Morgan who formulated a symbolic logic. De Morgan also showed that there are deductions that cannot be handled by the traditional Aristotelian logic. During the Logicism, Frege developed first-order predicate logic and tried to base all mathematics on mathematical logic. While mathematical logic was very successful for many years, Russell showed with his set paradox that Frege's logic contained contradictions.
During Hilbert's program mathematicians tried to ground mathematics on a finite and complete axiomatic system that is free of such contradictions. However, 1931 Gödel showed that in each rich enough and sound arithmetical theory there is a canonical and unprovable \textit{Gödel Sentence} which indirectly states that it is unprovable~\cite[p.171]{smithIntroductionGodelTheorems2020}. The Gödel Sentence is hence true and contradicts the goal of Hilbert's program.

Sentences that cannot be derived consistently as well as infinite regressions especially appear if a sentence contains self-references~\citep{kordig_self-reference_1983,bolander2002self}. As an example, let us refer to the statement: ``I am lying''. If the subject (I) is right, then it is lying and the statement is wrong, which would imply that it is not lying, and so on. Since paradoxes like the \textit{liar paradox} create unsolvable contradictions, Russell's student Wittgenstein argued that self-references should be banned from logical statements~\cite[3.332]{wittgensteinTractatusLogicoPhilosophicus2001}. However, this would also disallow circular but valid statements like: \textit{This statement has five words.}\footnote{A special case are sentences that contain ambiguous self-referring words like 'I' or 'Me'. Following \cite[p.5f]{wElementaryLogic1965}, declarative sentences like 'Ha' ('H' stands for 'is hungry' and 'a' stands for me) only become statements if the 'a' is supplanted by an unambiguous word (see also: \cite[p25]{gunther_idee_1991}). } Other approaches argue that self-referring statements are valid if one accounts for the context-dependence (common knowledge, previous arguments, ...) of the propositions. Hence, during the course of the paradox, the context might shift and thereby change the truth of the propositions \citep{parsonsLiarParadox1974,glanzbergLiarContext2001,burgeLiarParadoxTangles1982}. While the contextualist approaches might be valid explanations for the \textit{liar paradox}, they apply mainly to paradoxes that work on a semantic level and rely on the notion related to truth. However, there are also paradoxes like Russell's Set\footnote{Russell's Set denotes the set of all sets that are not members of themselves. The paradox occurs if one asks whether Russell's Set contains itself.} that operates on a syntactic level and demonstrated contradictions in Frege's First-order logic. In order to avoid this type of paradox the standard form of modern set theories, the Zermelo-Fraenkel set theory, axiomatically disallows Russell's set. A more versatile approach to tackle paradoxes was introduced by Tarski with the introduction of a hierarchy of language where truth predicates can only apply to sentences at lower levels which essentially excludes the used language from the domain of possible references \cite[p.350]{tarskiSemanticConceptionTruth1944}\cite[p.188]{gunther1964bewusstsein}. More radical solutions even suggest that statements like the \textit{liar paradox} do not have a truth value at all \citep{kripkeOutlineTheoryTruth1975}. This often implies the introduction of \textit{undefined} as a third truth-value and hence the rejection of the \textit{Tertium non datur} principle.

According to Günther, all known approaches to handle antinomies require a global subject (or a meta-language) that observes the world as an isolated object without interaction (referencing) to it \cite[p. 328]{gunther_beitrage_1976}. However, and in accordance with the observations of Fichte and Hegel, our world consists of endless self-referencing systems. An example from biology are gene-regulatory networks, where gene \textit{A} inhibits its own production. This interaction is not understandable if viewed statically with dimensionless topology diagrams~\citep{isalanGeneNetworksLiar2009}.\footnote{Of course, this creates only a logical paradox, if all interactions are considered to be instantaneously, which is not realistic. However, it is meaningless to search for an absolute answer, whether the gene is active on or inactive.} Also in social sciences self-referencing systems are more a rule than an exception: During a discussion, each member forges the opinion of the group the member itself belongs to \citep{klagenfurt_technologische_2016}.

Based on the assumption that truth is a subjective property that cannot be described with a two-valued logic\footnote{One might argue that modern predicate logic allows for the formulation of subjectivity by using predicates that have argument places for subjects, e.g. $X$ thinks, that $Y$ is wrong. However, in this case, 'We' as the one who replaces $X$ and $Y$ with actual values and who evaluates the truth of the statement are the actual subject, leaving $X$ and $Y$ as objects.}, Günther developed the theory of polycontexturality \citep{gunther_beitrage_1976}. This theory claims that reality has a polycontextural structure, which means that every interacting subject spans an isolated two-valued system, a \textit{contexture}, that has an inherent definition of True and False. In each contexture a binary logic is valid. Hence, Günther gives up the closed and viewpoint independent definition of a two-valued logic and generalises Hegel's and Fichte's viewpoint dependence to a new viewpoint dependent and multi-valued logic~\citep{bierter_wege_2018}. One should note that this multi-valued logic is not to be understood as a single multi-valued system as e.g. found in Fuzzy Logic~\cite[p.100]{bierter_wege_2018}\citep{zadeh_fuzzy_1965} or in the three-valued logic of Lukasiewicz and Kleene~\citep{lukasiewiczElementsMathematicalLogic1963, kleeneIntroductionMetamathematics1952,kleeneNotationOrdinalNumbers1938},
but rather as a network of interconnected two-valued systems, a place-value system (\textit{Stellenwert-Logik}), where objects and subjects have different place values. Due to this distributed definition of logical values, there is no objectivity but only subjective and contexture dependent interpretations and descriptions of measurements. In this new logic something can be true and false at the same time, depending on the contexture.

While Günther's theories are often utilised in the fields of sociology and cognition theory~\citep{jansen_who_2016,jansen_kontexturanalyse_nodate,jansen_beyond_2017,vogd_professions_2017,molders_differenzierung_2012,vogd_polykontexturalitat:_2013,gronoldKausalitaetGewaltKulturwissenschaftliche2014}, only little work was done to demonstrate the impact of Günther's ideas in the natural sciences~\citep{bruni_towards_2015}. At the same time, natural sciences like physics have been very successful in describing and explaining our world based on the classical two-valued logic. It is therefore interesting to analyze the potential role of contextures and polycontextural logic in physics. Here we show how Günther's Theory of Polycontexturality provides means to mathematically formalize observer-dependence, a problem that physics has already solved by transformations. Accordingly, it seems possible that our world is in fact polycontextural and our theories compensate for it.

Heisenberg stated: ``\textit{Natural science does not simply describe and explain nature; it is part of the interplay between nature and ourselves; it describes nature as exposed to our nature of questioning}''~\citep{heisenberg_physics_2000}. Here, Heisenberg makes clear that physical observations and the observer are always interwoven~\citep{bierter_wege_2018}, a problem that also puzzled physicist like Niels Bohr when they tried to interpret the results of quantum mechanics~\citep{mohanty1989idealism}. Since Einstein's theory of special relativity~\citep{einstein_zur_1905}, it became evident that even the perception of time is only relative and hence observer-dependent. As motivated in the previous paragraphs, view\-point-de\-pen\-dent truth does also imply that properties can apply and not apply at the same time. While this might sound unfamiliar at first, such phenomena exist even in modern physics e.g. the particle-wave duality~\citep{selleri_wave-particle_1992}: Depending on the measurement one can come to both the conclusion that photons are waves or particles. This contradicts a classical point of view where (again relying on Aristotelian logic) an object can only be particle or wave but not both at the same time. Using methods borrowed from Günther's theory we describe the observed phenomena in a polycontextural formalism. Subsequently, we discuss the question of why physics is able to describe natural occurrences even so it is not knowingly based on a multi-valued polycontextural logic. We observe that the necessary mapping between a polycontextural world and a two-valued science is only possible due to the utilisation of transformations. These transformations are used to convert an observer-dependent and hence subjective description of nature into the subjective position of another observer (another subject). 

The aim of our paper is twofold: On the one hand, we show how Gotthard Günther's concepts are (mostly inherently) already present in the current understanding of physical properties. On the other hand, we demonstrate that, given the assumption that our world is polycontextural, transformations are essentially an expedient that is used to make two-valued sciences compatible with it.

The remainder of this paper is organised as follows: In the next section (Section~\ref{sec:poly}), we introduce Gotthard Günther's concept of a polycontextural logic and briefly explain the so-called proemial relation, which forms the minimal model of a self-referring system that enables the notion of subjectivity. How Günther's theory can be formalised and hence be used to make dialectics operational is then explained in Section~\ref{sec:threevalued}. In Section~\ref{sec:time} we describe a small thought model that illustrates time dilatation and show in Section~\ref{sec:relpoly} how the concept of polycontexturality can be applied to subjective and in-congruent time measurements. Subsequently, in Sections~\ref{sec:phot} and~\ref{sec:dualpoly} we discuss the phenomenon of the particle-wave duality of photons and show how the concepts of the theory of polycontexturality fit into this understanding. In the last section (Section~\ref{sec:conclusion}), we draw some conclusions and motivate a further and more detailed discussion of the connections between modern physical and polycontextural theories.

\section{Theory of polycontexturality \label{sec:poly}}

\begin{figure}
    \centering
    \includegraphics{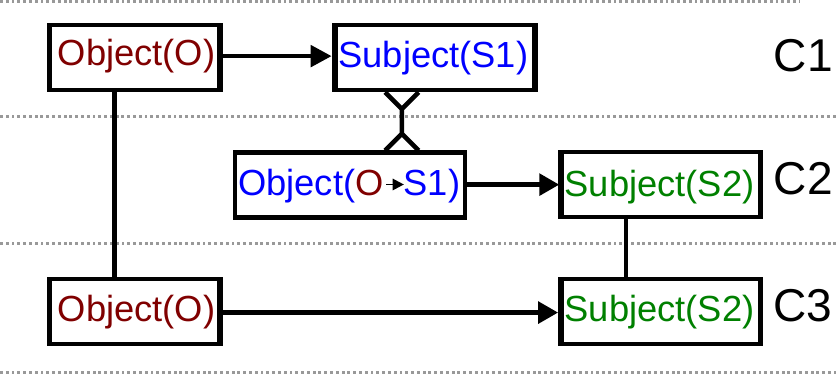}
    \caption{Sketch of the proemial relation. In each contexture (C1, C2, C3) a subject observes an object as indicated by the arrows (The direction of the arrows indicates the flow of information; Object $\rightarrow$ Subject reads: Subject observes Object). The lines (\textit{coincidence relations}) indicate,  that the red object in C1 and C3 and, respectively, the green subject in C2 and C3 are the same. The blue subject in C1 and the blue object in C2 are basically the same but change the role from an observing entity to an observed one (\textit{exchange relation}; indicated by the inverted double-arrow). The blue object in C2 represents the entire observation process of C1. (Figure adapted from:~\cite[p.49]{klagenfurt_technologische_2016}). }
    \label{fig:proemial_so}
\end{figure}

Günther postulated that reality consists of an unlimited number of so-called contextures. Each contexture has its own (and possibly unique) classical two-valued logic. Each of these values is a truth-value and, at the same time, indicates its place within a polycontextural system. Globally, this creates a multi-valued -- a place-valued -- logic of different subjects in different contextures. All contextures have equal rights and are aligned in a heterarchy (rather than a hierarchy). Hence there is no absolute subject as in Aristotelian logic. Using this concept, Günther noticed that the interaction of three contextures allows for the closed formulation of self-reference: He assumed that each subject and the objects it observes form a contexture, in which subject and object are connected by an \textit{order relation} that subordinates (in a static picture) the object to the subject. The process of a subject that observes an object in one contexture can, in turn, become object in another contexture (as sketched in Fig.~\ref{fig:proemial_so}). This exchange of an observing subject and an observed object is mediated by an \textit{exchange relation} that connects two contextures. In this way, in the second contexture, the initial object that was observed in the first contexture becomes an object as observed by the subject of the first contexture \citep{klagenfurt_technologische_2016}. It is necessary to stress again that contextures form a heterarchical order. Hence, the subject of the second contexture can also become the subject of the first contexture that is observed by some other subject. In contrast to propositional logic, no hierarchy puts one contexture a level above another contexture~\cite[p.228]{gunther_beitrage_1979}. 
Finally, a third contexture mediates the relationship between the original object of investigation and the same object in its subjectified form. The fact that the object of the first contexture is also the object of the third contexture, as well as that the subject of the third contexture is the subject of the second contexture is reflected by \textit{coincidence relations} (lines in Fig.~\ref{fig:proemial_so}). The third contexture (that is only created by the intertwining interaction of the first and second contexture) hence provides the means for the individual subjects to reflect their own understandings of the world. Günther termed this structure of three contextures \textit{proemial relation}.

By naming the three different processes, Günther makes the structures and their respective functions more explicit. The internal subject of the first contexture observes an object. If this internal subject becomes an object of another external observer, this is formally equivalent to an \textit{I} that thinks about a \textit{You} whereby both the \textit{I} and the \textit{You} are placed in the same environment. Obviously, a simple change of the perspective interchanges the \textit{I} and the \textit{You}, which highlights the heterarchical character of both contextures. Both the \textit{I} and the \textit{You} are objects with subjective perception and the \textit{I} is the \textit{You}'s you. In the mind of the \textit{I} the \textit{You} exists and likewise, the \textit{I} exists in the mind of the \textit{You}. Nevertheless, \textit{I} and \textit{You} are completely different from each other. There is no global perspective that subsumes both the \textit{I} and the \textit{You}~\citep{botz-bornstein_i_nodate}.  Let us now assume that the \textit{I} and the \textit{You} are talking about the same topic, the \textit{It}. For the \textit{I} this \textit{It} is now accessible both directly or via the observation of the \textit{You} that talks about the \textit{It} (the \textit{You} objectified the \textit{It}). This mediation between two perspectives creates the third contexture and likewise indicates the \textit{I} how subjectivity manifests in \textit{You}'s observations (based on the \textit{I}'s subjective world view)~\cite[p.80]{gunther1964bewusstsein}.

In the Aristotelian logic, this construction must fail. If the \textit{You} is thought of by the \textit{I} then the \textit{You} necessarily needs to be a simple ``dead object'' and cannot be thought of as a thinking subject. Here, the designation ``dead object'' only refers to the standpoint of the \textit{I}. It is, of course, possible that the \textit{I} thinks about a \textit{You} thinking about dog. But in doing so, the \textit{I} thinks in its own logical context throughout. The \textit{I} can't think about what the \textit{You} is thinking about in the \textit{You}'s context/using the \textit{You}'s logical system. As soon as one agrees that \textit{I} and \textit{You} are different and \textit{I} and \textit{You} perceive differently, a globally valid logic must fail and one needs to take polycontextural logic into account. Each subject and its perception span a defined contexture and can be described with a two-valued logic. Nevertheless, the totality of all subjects calls for a higher-valued, a polycontextural logic~\cite[p. 87]{gunther1980beitrage}.

\begin{figure}
    \centering
    \includegraphics[width=.5\textwidth]{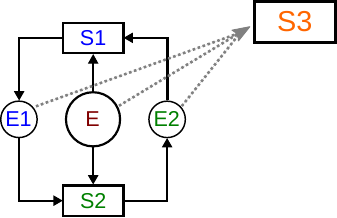}
    \caption{Sketch of two subjects $S1$ and $S2$ observing the same environment $E$. Both subjects interact with the environment and produce events $E1$ and $E2$ that contain their subjective information about the observed environment. A third subject $S3$ can observe the convergence process of $E1$ and $E2$. (Figure adapted from:~\cite[p.88]{gunther1980beitrage}).}
    \label{fig:subject_environment_circle}
\end{figure}

\begin{figure}
    \centering
    \includegraphics[width=\textwidth]{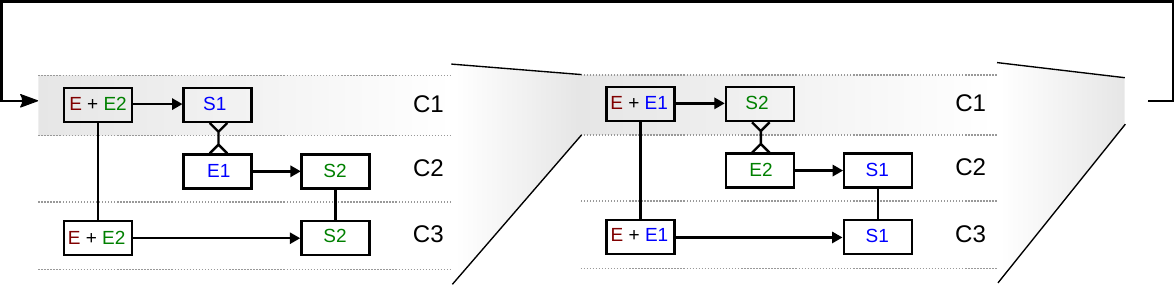}
    \caption{Two subjects and the environment form a proemial relation. This proemial relation itself is a contexture of another relation which creates a cyclic structure of two connected relations.}
    \label{fig:subject_environment_connected}
\end{figure}

Following \cite[p.88]{gunther1980beitrage}, we illustrate this concept with a small example. In order to explain how two interacting subjects form a mutual \textit{I}-\textit{You} system that is unable to generate absolute objectivity, we draw on Fig.~\ref{fig:subject_environment_circle}. Here, $E$ is the environment that is observed by two subjects $S1$ and $S2$, where both subjects are assumed to perceive different aspects of $E$. The information flow from $E$ to the respective subject is indicated with an arrow. From the perspective of each subject the environment contains one other subject that itself observes the environment and interacts with it. The interactions of $S1$ and $S2$ with the environment generate additional information about the environment that are denoted $E1$ and $E2$, respectively. As a result, $S1$ not only observes $E$ but also $E2$ generated by $S2$ and likewise $S2$ observes $E$ and $E1$. Hence, although $S1$ and $S2$ are embedded in the same environment their respective observations are different. It is reasonable to assume that the events $E1$ and $E2$ are somehow related to $E$. Therefore, $E1$ can be considered to contain $S1$'s subjective information about $E$ (the same holds for $S2$ and $E2$). As a result, an updated $E1$ will not only depend on $E$ but also on $E2$ which itself depends on $E1$. Over the course of time, both subjects will agree on a common description for the same observed features and the subjective understandings $E1$ and $E2$ will converge. Nevertheless, since we can only distinguish between $S1$ and $S2$ based on the difference between $E1$ and $E2$, they will never become equal and there will never be a common and hence objective understanding of $E$.

The described mutual observation of the two interacting subjects can also be sketched as two co-existing proemial relations, as shown in Fig.~\ref{fig:subject_environment_connected}. One proemial relation is formed by $S2$ that observes both $E1$ and $E$. This first relation can hence be written as $E+E1 \rightarrow S2$, which itself is part of another relation. This second relation can be condensed to $E+E2 \rightarrow S1$, which is again an integral part of the first contexture. It is important to note the relations do not represent consecutive but coexisting states of the system.

We can now imagine that there is a third subject $S3$ observing the full structure of the two interacting subjects $S1$ and $S2$. For $S3$ both $S1$ and $S2$ belong to the objective world. Hence, $S3$ can observe the convergence process of $E1$ and $E2$ as an uninvolved individual. Additionally, $S3$ is able to understand that $S1$ and $S2$ are both right in what they preceive and reveal about $E$, even so, $E1 \neq E2$. It is, however, important to note that the existence of $S3$ does not imply that $S1$ and $S2$ can be described from an objective viewpoint. Rather is $S3$ itself a subject of the shown setup and can, e.g., adopt the position of $S1$, which can then be observed by $S2$. Likewise $S1$ or $S2$ can take up the position of a spectator observing $S3$ and the respective other subject. Hence, all of the three subjects are equal in what observer position they can take up: they form a heterarchy. Nevertheless, as long as we assume that $S3$ acts as an external spectator, and being aware that this implies a freezing of the in reality perpetually changing roles, we as readers ourselves can take up the position of the third subject. Hence, in the following examples we will not specifically mention the third subject but always assume that the reader takes up this position.

With the arguments presented so far, Günther justifies why a polycontextural description is suitable for representing real physical relationships that are excluded in binary systems. So far, this theory is in large parts a reiteration of the concepts of modern dialectic mainly developed by Hegel. However, by the introduction of a three-valued logic, Günther was able to make dialectics operational. In the following section we will outline the concept of this three-valued logic and subsequently apply it to two examples taken from physics.

\section{Three-valued logic}
\label{sec:threevalued}
The \textit{Leibniz law}, one of the most fundamental laws in logic states, that two entities A and B are equal if A has every property of B and vice versa~\citep{tarskiIntroductionLogicMethodology1995}. Hence, every entity is only identical to itself. If the existence of such a self-identical entity is stated by a positive predicate, then -- due to the law of noncontradiction -- the negation of this predicate can only exist in a reflection. Now, the law of excluded middle asserts that there can not exist a third between the given positive and negative predicates. This indicates that classical logic requires a direct symmetry relation between Being and Thought~\cite[p.127]{gunther_idee_1991}. As a consequence, in classical binary logic everything that is not a subject is an object and everything that is not an object is necessarily a subject.

This contradicts the ideas of the polycontextural concept described above where different observers (different subjects) are able to reflect objects as well as other reflecting observers. Günther acknowledged that binary logic is the correct way for every single subject (working in its personal contexture) to describe its observations. However, he argued that classical logic is only local and intra-subjective.

\begin{figure}
    \centering
    \includegraphics[width=\textwidth]{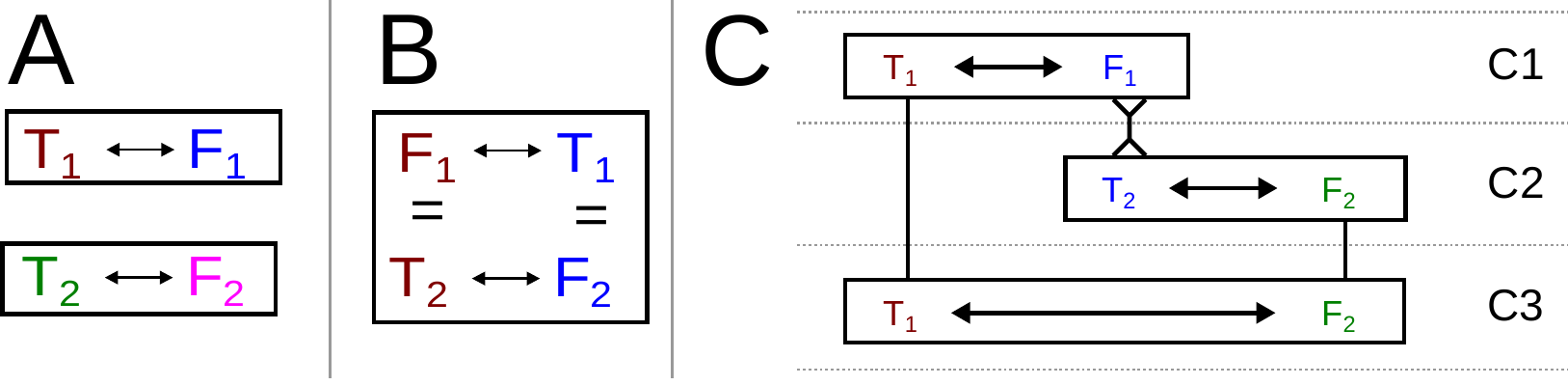}
    \caption{Sketch of the three examples presented in the main text. A: Two contextures with isolated logical systems. B: Two subjects in one contexture with a direct mapping between $T_1, F_1, T_2, F_2$. C: Three contextures that form a proemial relation (see also Fig.~\ref{fig:proemial_so}).}
    \label{fig:proemial_truth}
\end{figure}

Let us assume two different subjects $S_1$ and $S_2$, each equipped with its own binary logical values $T_1 \leftrightarrow F_1$ and $T_2 \leftrightarrow F_2$. If both subjects do not interact we arrive at a four-valued logic with two completely isolated contextures (see also Fig.~\ref{fig:proemial_truth}A). Since they do not exchange information, both subjects have no means to learn whether their logics can be mapped on each other or not. Contrary, if we assume that there is a direct mapping between $T_1 \leftrightarrow F_1$ and $T_2 \leftrightarrow F_2$, e.g. $T_1 = F_2$ and $T_2 = F_1$, both subjects are equal or just the mirror-image of each other (Fig.~\ref{fig:proemial_truth}B). 

Now, let us assume that there is only a mapping between one of the respective logical values, e.g.: $F_1 = T_2$ but $T_1 \neq F_2$. In a classical understanding this would lead to a contradiction, since
\begin{equation}
    \lnot T_1 = F_1 = T_2 = \lnot F_2 \rightarrow \lnot T_1 = \lnot F_2 \rightarrow T_1 = F_2~.
\end{equation}
However, Günther argues that each contexture has its own negation, leading to 
\begin{equation}
   \lnot_1 T_1 = F_1 = T_2 = \lnot_2 F_2  \rightarrow  \lnot_1 T_1 = \lnot_2 F_2.
\end{equation}
The problem of how $\lnot_1$ and $\lnot_2$ are related can then be solved by a third subject $S_3$ that spans another contexture with the logical values $T_1 \leftrightarrow F_2$. Figure~\ref{fig:proemial_truth}C shows how this concept relates to the proemial relation explained in the previous section. Hence, the object of the second contexture is not an irreflexive object but the reflection process of the first contexture and the negation $\lnot_2$ negates the complete alternative situation of C1.

Obviously, this concept renders the global assignment of True and False to specific logical values impossible. Similarly to Günther we will use the integer values $1\equiv T_1, 2\equiv F_1/T_2, 3 \equiv F_2$ instead of $T$ and $F$ from now on. To understand Günther's concept, it is important to note that these values are not numbers in the conventional sense, but place-values. Two values each define a logical place in which both assume the function of truth-values. Hence, each of these place-values is a truth-value and, at the same time, indicates its place within a polycontextural system.

In order to operationalize this framework, Günther showed how the standard logical functions like AND and OR can be understood in this polycontextural system. In Tab.~\ref{tab:binaryandor} we show the truth tables for the case of the standard binary logic, where we relabeled $T \rightarrow 1, F \rightarrow 2$. A mechanistic definition of the two logical functions could hence be that AND (OR) always takes the largest (smallest) value of the two given alternatives. This definition can be used to generalize both logical functions for a polycontextural logic of three contextures as shown in Table~\ref{tab:polyandor}. We then arrive at a definition of a multi-valued logic that -- in its structure -- is known from the works of Lukasiewicz~\citep{lukasiewiczElementsMathematicalLogic1963} or Kleene~\citep{kleeneIntroductionMetamathematics1952,kleeneNotationOrdinalNumbers1938}. 

\begin{table}
    \centering
\begin{tabular}{c|c||c}
    p & q & p $\land$ q \\ \hline
    1 & 1 & 1  \\
    1 & 2 & 2  \\
    2 & 1 & 2  \\
    2 & 2 & 2 
\end{tabular}
\quad
\begin{tabular}{c|c||c}
    p & q & p $\lor$ q \\ \hline
    1 & 1 & 1  \\
    1 & 2 & 1  \\
    2 & 1 & 1  \\
    2 & 2 & 2 
\end{tabular}
\caption{Logical AND (left) and logical OR (right)in classical binary logic. Here, 1 denotes the classical T and 2 the classical F.}
\label{tab:binaryandor}
\end{table}

\begin{table}
\centering
\begin{tabular}{c||c c||c || c  }
        & p & q & p $\land$ q & $p \lor q$   \\ \hline
    1   & 1     & 1 & 1 & 1    \\
    2   & 1     & 2 & 2 & 1   \\
    3   & 1     & 3 & 3 & 1   \\
    4   & 2     & 1 & 2 & 1   \\
    5   & 2     & 2 & 2 & 2   \\
    6   & 2     & 3 & 3 & 2    \\
    7   & 3     & 1 & 3 & 1  \\
    8   & 3     & 2 & 3 & 2    \\
    9   & 3     & 3 & 3 & 3  \\ \hline \hline 
    & & & & 
\end{tabular}
\quad
\begin{tabular}{c||c c||c c c | }
        & p & q &  \multicolumn{3}{c}{p $\land^D$ q}   \\ \hline
    1   & 1     & 1     & 1     &   & 1 \\
    2   & 1     & 2     & 2     &   &    \\
    3   & 1     & 3     &       &   & 3  \\
    4   & 2     & 1     & 2     &   &    \\
    5   & 2     & 2     & 2     & 2 &    \\
    6   & 2     & 3     &       & 3 &    \\
    7   & 3     & 1     &       &   & 3  \\
    8   & 3     & 2     &       & 3 &    \\
    9   & 3     & 3     &       & 3 & 3  \\ \hline \hline
        &   &   & S1 & S2 & S3  
\end{tabular}
\caption{Left: logical AND and logical OR for a polycontextural logic with three connected contextures. Right: rearranged logical AND with a separate column for each contexture (S1,S2,S3). For each contexture only the relevant positions are filled in.}
\label{tab:polyandor}
\end{table}

We will now analyze how Günther extends this logic in order to meet the requirements of an operational dialectic. 
In the following, we will focus on the logical AND, however, the results can analogously be applied to the logical OR. 

Given the three contextures, each based on a binary logic, we can arrange the results of Tab.~\ref{tab:polyandor} (left) in such a way that each contexture is written in a separate column, as depicted in Tab~\ref{tab:polyandor} (right). Due to our initial global definition of the logical function the contextures S1 and S2, both show a logical AND and are framed by an overarching AND of the third contexture (S3). For this reason, Günther calls this function a \textit{total conjunction} (given by the operator $\land^D$) and contrasts it by two \textit{partial conjunctions} (given by the operators $\land^R$ and $\land^I$) that are distinguished by a disjunction in either S1 or S2 as depicted in Tab.~\ref{tab:polyfulland}. A more detailed interpretation of the tables will be presented in the following examples.

\begin{table}[h]
\centering
\begin{tabular}{c||c c||c c c | c c c | c c c }
        & p & q &  \multicolumn{3}{c}{p $\land^D$ q} & \multicolumn{3}{c}{p $\land^R$ q} & \multicolumn{3}{c}{p $\land^I$ q}  \\ \hline
    1   & 1     & 1     & 1     &   & 1 & 1 &   & 1 & 1 &   & 1 \\
    2   & 1     & 2     & 2     &   &   & 1 &   &   & 2 &   &   \\
    3   & 1     & 3     &       &   & 3 &   &   & 3 &   &   & 3 \\
    4   & 2     & 1     & 2     &   &   & 1 &   &   & 2 &   &   \\
    5   & 2     & 2     & 2     & 2 &   & 2 & 2 &   & 2 & 2 &   \\
    6   & 2     & 3     &       & 3 &   &   & 3 &   &   & 2 &   \\
    7   & 3     & 1     &       &   & 3 &   &   & 3 &   &   & 3 \\
    8   & 3     & 2     &       & 3 &   &   & 3 &   &   & 2 &   \\
    9   & 3     & 3     &       & 3 & 3 &   & 3 & 3 &   & 3 & 3 \\ \hline \hline
        &   &   & \color{gray}S1 & \color{gray}S2 & \color{gray}S3 & \color{gray}S1 & \color{gray}S2 & \color{gray}S3 & \color{gray}S1 & \color{gray}S2 & \color{gray}S3 \\
      &   &   & \multicolumn{3}{c}{O3} & \multicolumn{3}{c}{O2} & \multicolumn{3}{c}{O1}
\end{tabular}
\caption{Complete polycontextural table for the three possible AND functions. Here, $p \land^D q$ denotes the \textit{total conjunction}; $p \land^R q$ and $p \land^I q$ are the two \textit{partial conjunctions}.}
\label{tab:polyfulland}
\end{table}

In the next section, we will introduce a thought experiment that presents an apparent situation of observer dependence in physics. Using the ideas of polycontextural logic we then show how the incompatibility of different measurements can be incorporated into the formalism, consequently becomes an integral part of the formalism, and, thus also of our everyday intuition about this class of physical systems. Subsequently, we analyse how modern physics utilises transformations to address and dissolve polycontexturality in these situations.

\section{Relative moving reference systems as contextures}
\label{sec:time}

We refer to an example from the theory of special relativity. This theory, proposed by Albert Einstein in 1905, is one of the cornerstones of theoretical physics. The two basic postulates of this theory are~\citep{stachel_collected_1990}: 
\begin{itemize}
    \item The laws of physics are invariant in all reference frames.
    \item The speed of light is the same for all observers.
\end{itemize}
\begin{figure}
    \centering
    \includegraphics[width=\textwidth]{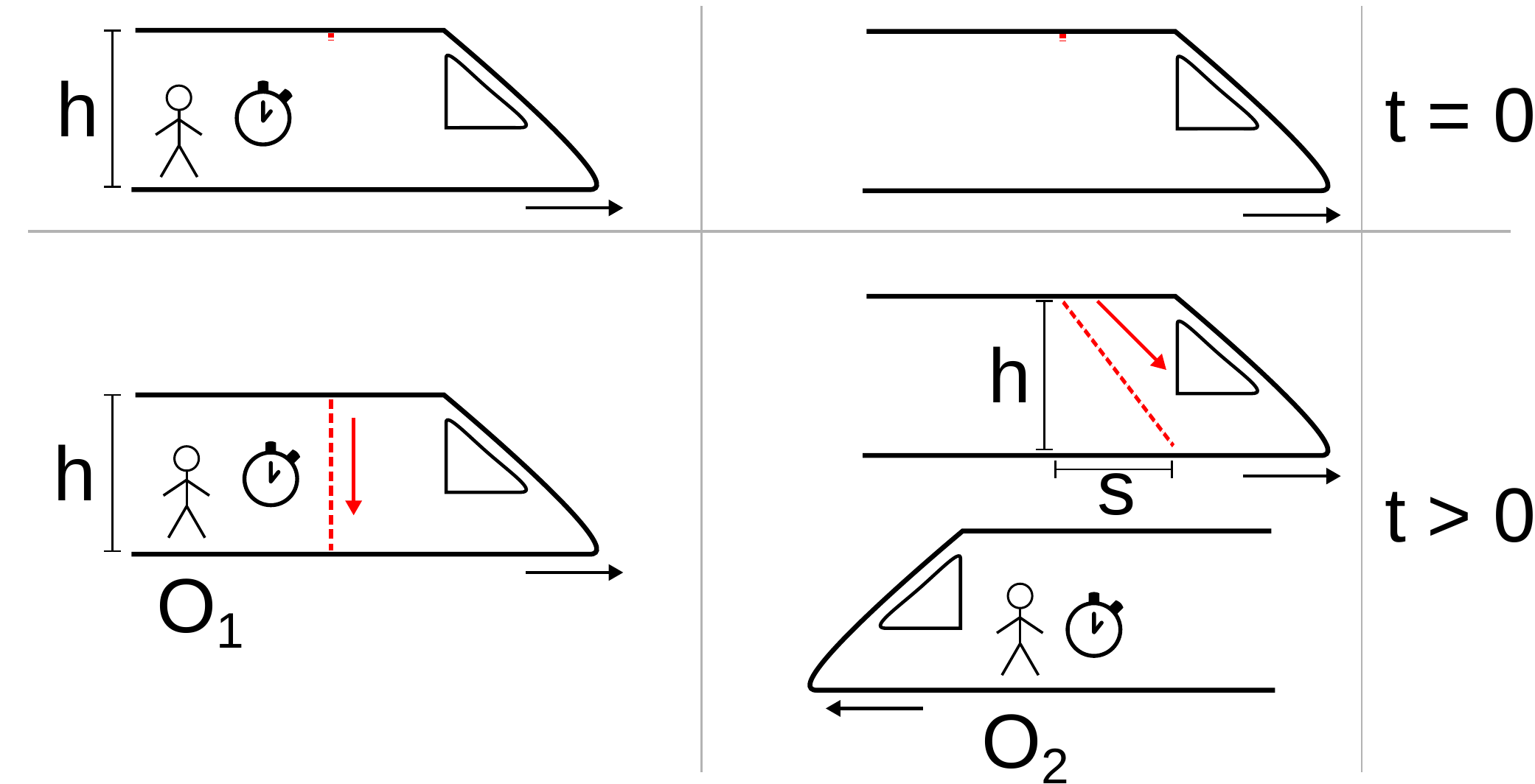}
    \caption{Sketch of the thought model to explain time dilatation. (left) One observer $O_1$ inside the train observes the laser-beam. (right) The observer $O_2$ observes the beam from the perspective of another train.}
    \label{fig:train}
\end{figure}
While these postulates are well known, their implications contradict daily experiences and, furthermore, give rise to an observer dependence of some measured quantities. In what follows we focus on one special implication of the above postulates, namely the relativity of time. Let us assume two observers $O_1$ and $O_2$ seated in two similar trains (as depicted in Fig. \ref{fig:train}). The trains shall pass each other at a relative speed of $v = 0.5 c$, where $c$ denotes the speed of light. This might e.g. be due to both trains having an opposite speed of $v' = 0.25 c$ or similarly due to one train standing still and the other one passing at a speed of $v' = 0.5 c$. 
Let us further assume that observer $O_1$ measures in its train the time a laser beam needs to go perpendicularly from the ceiling down to the floor (Fig.~\ref{fig:train}~left). Given the speed of light $c$ and the height of the train $h$ we can predict the exact travel time $t_1$ that observer $O_1$ obtains:
\begin{equation}
t_1 = \frac{h}{c}\,.
\end{equation}
If, however, $O_2$ wants to do the same measurement on the laser beam in $O_1$'s train, things are different. During the time interval $t_1$ the first train moves a distance of $s = v t_1$ relative to the second train. Due to this relative movement, $O_2$ will not observe a perpendicular but a diagonal trajectory of the laser beam that now goes down the height of the train $h$ but at the same time moves forward a distance of $s$ (Fig.~\ref{fig:train}~right). Hence, from $O_2$'s point of view, the laser beam travels a distance of:
\begin{equation}
    h' = \sqrt{h^2 + s^2}\,.
\end{equation}
One of Einstein's postulates states that the speed of light is constant. Thus, the observer $O_2$ obtains a travel-time for the laser beam of:
\begin{equation}
    t_2 = \frac{h'}{c} = \frac{\sqrt{h^2 + s^2}}{c} < t_1\,,
\end{equation}

Both observers have conducted a measurement that fulfils all requirements of an objective and authoritative experiment that was -- in each train -- based on the valid laws of physics. However, if the observers meet and compare their measured times, they do not agree. Furthermore, there is no way to tell whose results are correct. Both moving trains constitute a reference frame where the laws of physics are invariant and hold. One can consider each train as a distinct contexture where a classical logical system (represented by the laws of classical physics) is valid.\footnote{Note that the comparison between classical logic and classical physics (no quantum mechanics and no theory of relativity) is more than a play on words. If the speed of the objects under investigation is much smaller than the speed of light, the laws of classical physics are a sufficient approximation. Analogously, as long as the interactions of the objects under investigation do not depend on subject-object relations, the laws of classical logic are sufficient.} This brings us back to Gotthard Günther's theory of polycontexturality.

\section{Lorentz transformation as a relation among contextures}
\label{sec:relpoly}
Using the ideas of Günther's theory of polycontexturality, we are in the position to interpret the phenomenon observed in the previous thought experiment as a result of observer dependence. Each observer performs a local measurement of the time the laser beam needs to pass the train. For $O_1$ inside the train all physical laws are valid and -- for $O_1$ -- the measurement seems to be an objective and universal measurement. Hence, $O_1$ and the observed laser beam form one contexture that exists inside the first moving train. The same holds for the laser beam and $O_2$. From the second train $O_2$ could perform an equally valid and universal measurement that, however, results in a different value for the time. In a mono-contextural interpretation (a global two-valued logic) this would create a contradiction.

Based on the ideas of the theory of polycontexturality, we can now assume that $O_2$ observes $O_1$ while $O_1$ is performing its measurement.\footnote{Analogously, the same can happen for $O_1$, who observes $O_2$.} In other words: $O_2$ observes the former observer $O_1$ while the latter observes the laser. Hence, $O_2$ observes now two objects: the laser and the laser as subjectively observed by $O_1$. As we already know, both observations will result in different values for the time.\footnote{As a result, $O_2$ might decide that $O_1$'s stopwatch is moving too slowly. This is not helpful in our current debate, since $O_1$ and $O_2$ still disagree in their results. We hence assume that both observers can be certain that the measurement was done correctly and the watches are perfect.} However, from a polycontextural view $O_2$ and the laser as well as $O_2$ and the former subject $O_1$ span two different contextures with different two-valued logics. It is hence up to $O_2$ to combine the two observations and communicate the resulting time $t_2$. In the same way, $O_1$ can directly measure the time of the laser as well as observe the result $t_2$ of $O_2$'s measurement (which itself depends on a previous result of $O_1$). During several such communication cycles between $O_1$ and $O_2$ (in which also the relative speed between the two observers might change), both observers will understand that the time measurement is a relative process and will eventually try to converge to a description of the laser that seems to be general for both observers. Since both observers are directly involved in this circular process, only we as an outside observer are able to objectively describe the convergence process.

While we are able to interpret and accept the validity of both divergent measurements using the theory of polycontexturality, the concept of the relativity of time is also well known and perfectly understood in classical physics. In the previous example, we explained that the two observers will converge on a description that seems to be general for both of them. If our assumption of a polycontextural world is correct this would imply that there is some transformation that enables one observer to map the subjective and polycontextural observation onto the own frame of reference. 

Indeed, due to the theory of special relativity we know that the relation that connects the observed subjective time $t_1$ of $O_1$ and the observable time is the Lorentz transformation:
\begin{equation}
    t' = \gamma \, t \,,
\end{equation}
where $\gamma = \frac{1}{\sqrt{1- v^2/c^2}}$.
Hence, for two observers with different speed, it is and will never be possible to define an absolute and objective time. The reference frames form a heterarchy.

This is connected to section~\ref{sec:intro} were we noted that classical logic does only allow for two values, namely \textit{True} and \textit{False}. Here a transformation at the core of the theoretical framework implements the relation among different contextures that allows for constructions like ``True in contexture 1 and at the same time False in contexture 2''.

 \section{Logically resolving observer-dependent measurements within a polycontextural framework}

We can also explain the described situation by the means of the logic tables presented in Tab.~\ref{tab:polyfulland}. For this, we identify the AND of the first observer $O_1$ with $p\land^I q$ and that of the second observer $O_2$ with $p \land^R q$. The first observer claims that the correct time is $t_1$, the second observer $O_2$ claims that the correct time is $t_2$. The two operands $p$ and $q$ can then be interpreted as:
 \begin{eqnarray}
     p &= \textrm{My measurement confirms my claim,} \nonumber \\
     q &= \textrm{Your measurement confirms my claim}. \nonumber
 \end{eqnarray} 
 With these assignments the logical AND in S1/O1 has the intuitive meaning: \textit{I consider my claim $t_1$ only as correct if my measurement AND your measurement confirm my claim.} 

As explained in the previous section, S2/O1 discusses the antithesis of the  (short: anti-claim): $\lnot t_1$. The meaning behind S2/O1 is hence: \textit{I consider my anti-claim $\lnot t_1$ as correct if my measurement OR your measurement confirm my anti-claim.}

 In classical logic this would just be a version of De Morgan's law, since: $2 \textrm{ in } S1 \equiv F \rightarrow 2 \textrm{ in } S2 \equiv T$ and equally $1 \textrm{ in } S1 \equiv T \rightarrow 3 \textrm{ in } S2 \equiv F$. However, within a polycontextural interpretation the negation $\lnot_2$ in the second contexture does not mirror back to the first contexture, but creates a third value that lies ``beyond'' the classical binary values. Hence, the third contexture S3/O1 discusses the relation between the claim and the anti-claim: $t_1 \leftrightarrow \lnot t_1$. 

 The same applies to the second observer $O_2$ that also spans three contextures. However, without loss of generality, we assume that for the second observer the claim $t_2$ is discussed in S2 and the anti-claim $\lnot t_2$ is discussed in S1. 

\begin{figure}
     \centering
     \includegraphics[width=\textwidth]{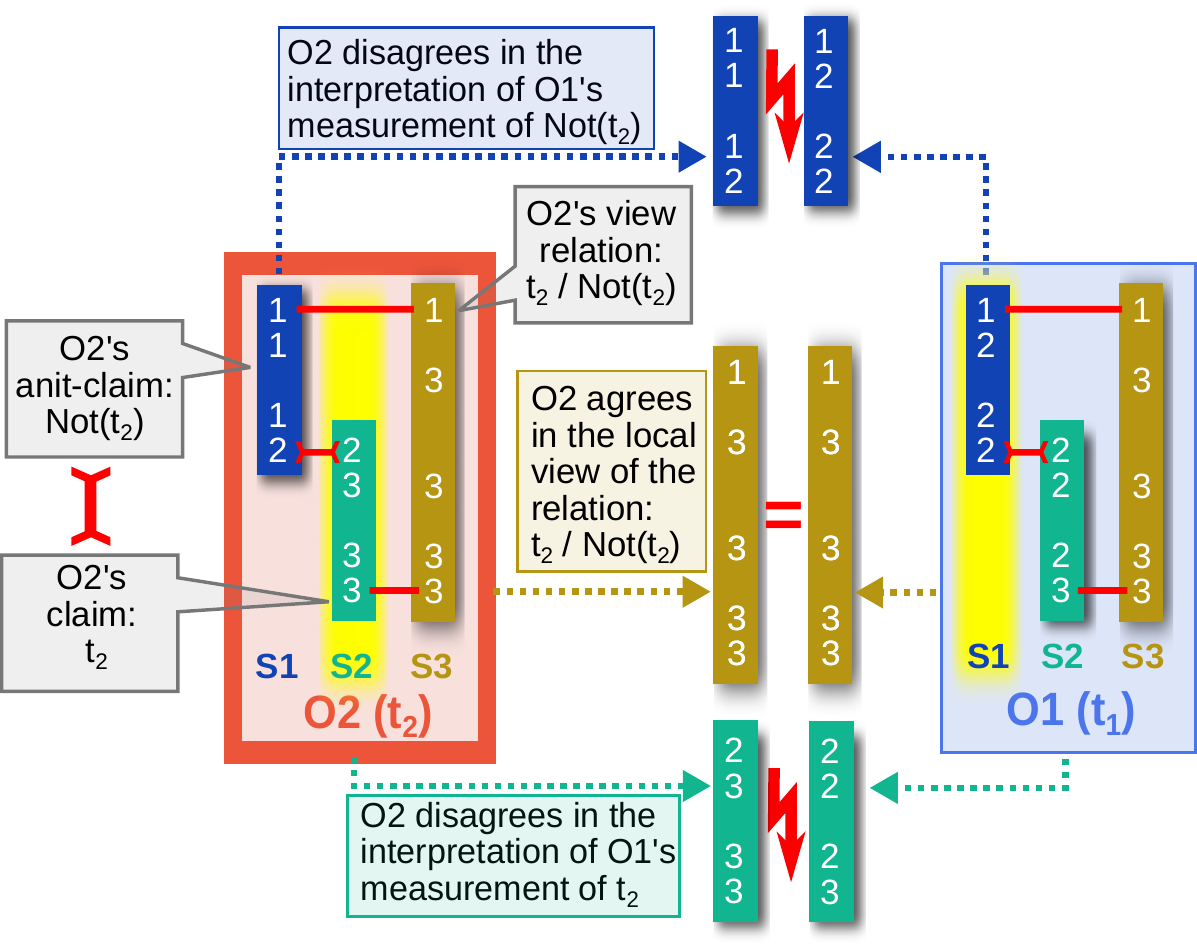}
     \caption{Sketch of the relation between O1 and O2 based on the local view of O2 with its claim $t_2$.}
     \label{fig:graphtwo}
 \end{figure}

 Up to now, both observers performed only a local discussion of their measurement. 
 If, however, O1 wants to compare its claim with O2, this is only possible via a comparison between $t_1$ and $\lnot t_2$, hence a comparison between an AND in S1/O1 and an OR in S1/O2. For O1 the only possible decision is hence, that O2's result is wrong. The same applies in the inverse direction to O2 who might compare his claim $t_2$ with O1 $\lnot t_1$. A graphical representation of these processes is depicted in Fig.~\ref{fig:graphtwo}. The unsolvable contradiction between $O1$ and $O2$ illustrates the ``unsolvable'' situation that we already described in the previous paragraphs. What is missing is a third observer.

\begin{figure}
     \centering
     \includegraphics[width=\textwidth]{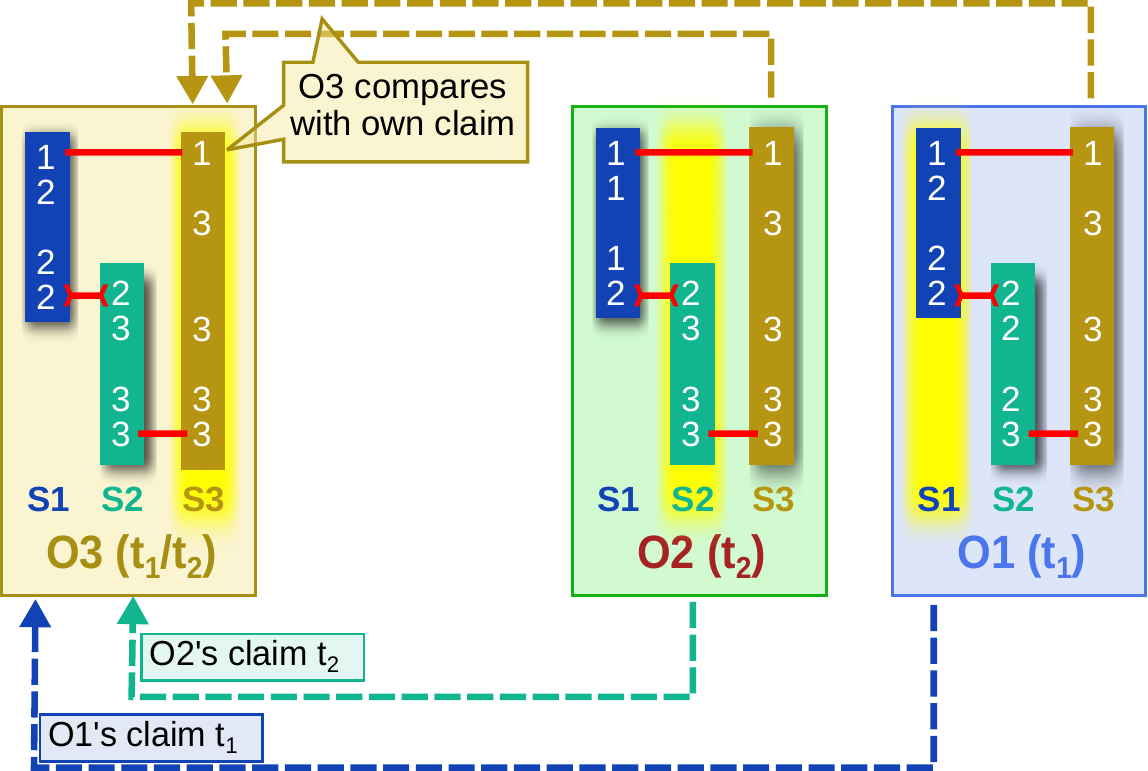}
     \caption{Sketch of the proemial-relation between O1, O2 and O3 based on the local view of O3 with its claim $t_1 \land t_2$.}
     \label{fig:graphall}
 \end{figure}

As a third and external observer we are in a different contexture, indicated in Tab.~\ref{tab:polyfulland} by $p \land^D q$. We can observe a \textit{difference} between O1 and O2 but, at the same time, we observe that the structures of their (mutual) observations are \textit{equal}. If we, therefore, compare S1/O1 and S2/O2 we are comparing two logical AND functions (they are now S1/O3 and S2/O3). As an external observer we can evaluate the correctness of both claims from the same (our) standpoint and need to acknowledge them. We need to conclude: \textit{Both are right}. Hence, the relation between both of these measurements needs to be a logical AND. A graphical representation of these processes is depicted in Fig.~\ref{fig:graphall}. 

 This ultimately leads to a logically closed description of the result that $t_1$ AND $t_2$ are both true, although different: The only logical conclusion for O3 is that O1 an O2 each have their own time system which is different from that of the other.

\section{The wave and particle properties of photons as distinct contextures}
\label{sec:phot}
A prominent but likewise confusing concept in quantum mechanics is the wave-particle duality, which states that particles like photons have the properties of a wave as well as of a particle. In the following, we describe two historical experiments illustrating this duality.

In 1802, the British physicist Thomas Young presented an experiment, which proved that light displays characteristics of waves. In this double-slit experiment, coherent light is shed on two parallel slits. The light that passes the slits causes a diffraction pattern on a screen. This pattern can only be explained if one assumes that light is a wave originating from the two slits and interfering on the screen. Using the geometry of the experiment as well as the distance of the observed maxima of the diffraction pattern it is possible to obtain the exact wavelength $\lambda$ of the light.

While the double-slit experiment suggests that light needs to be understood as waves, there are also experiments that seem to postulate an opposite understanding. In 1839 the French physicist Alexandre Edmond Becquerel observed the emission of electrons when light shines on materials. Interestingly, upon an increase of the light intensity the energy of the emitted electrons does not change but their number. This behaviour can only be understood if light is considered as a beam of small particles.

Hence, the photoelectric effect leads to the conclusion that light should be considered as particles instead of waves. Based on experimental observations it is possible to calculate the energy of one of such light particles, the photons:
\begin{equation}
    E = h f = h \frac{c}{\lambda}
\end{equation}
where $h$ is the Planck's constant, $f$ is the frequency of the photon and $\lambda$ is the wavelength of the light beam (as e.g. measured in the double-slit experiment). Using the mass–energy equivalence $E = m c^2$ one can also derive the momentum of a photon:
\begin{equation}
    E = m c^2 = p c = \frac{h}{\lambda} c \\
    \Rightarrow p = \frac{h}{\lambda}\,.
\end{equation}

These experiments and their implications are well proven and validated. Nevertheless, they seem to propose two contradicting concepts, namely that light displays characteristics of waves as well as characteristics of particles. Furthermore, none of both characteristics can be used to explain both experimental results. Hence, there is no possibility to decide on one correct interpretation. In fact, the common understanding of modern physics is that light shows a particle-wave duality, which means light is simultaneously a particle and a wave. Or to say it with Einstein's words: ``\textit{It seems as though we must use sometimes the one theory and sometimes the other, while at times we may use either. We are faced with a new kind of difficulty. We have two contradictory pictures of reality; separately neither of them fully explains the phenomena of light, but together they do}''~\citep{brown_evolution_1939}. This -- again -- brings us back to Günther's theory of polycontexturality.

\section{Wave–particle duality as a polycontextural phenomenon}
\label{sec:dualpoly}

\begin{figure}
    \centering
    \includegraphics[width=\textwidth]{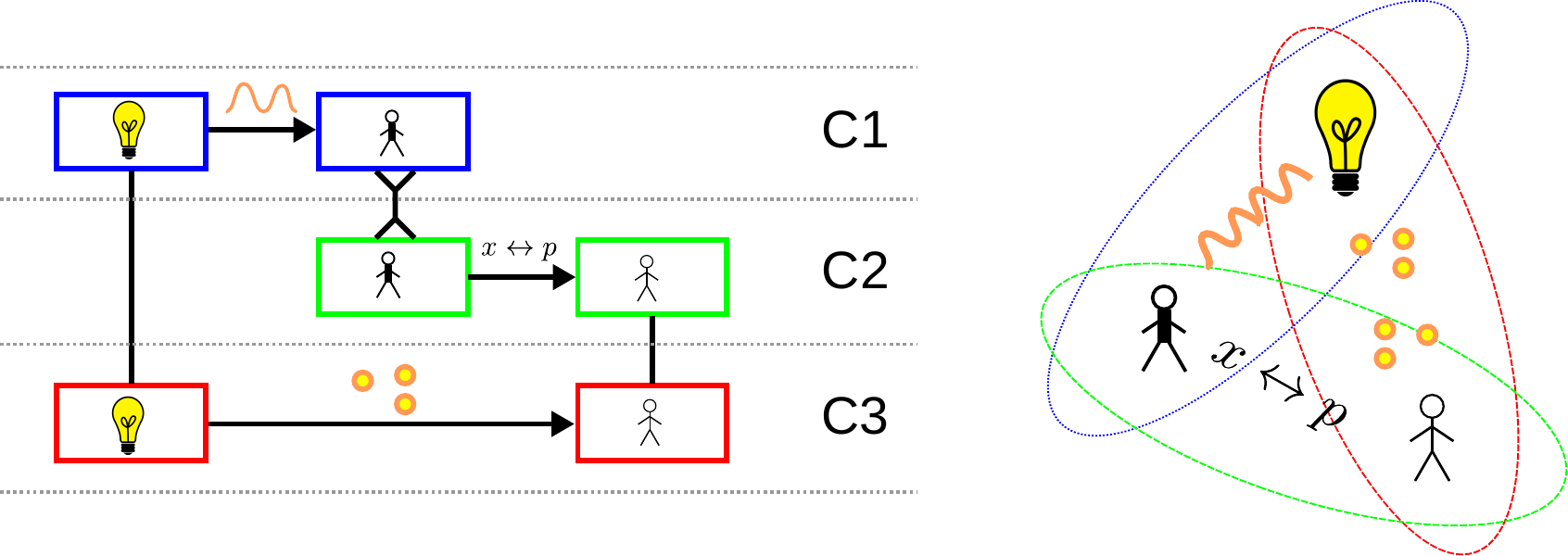}
    \caption{Sketch of the polycontextural analysis of the particle-wave duality. One observer detects particles, the other observer identifies waves.}
    \label{fig:duality}
\end{figure}

The two presented experiments -- and there are plenty more -- give contradicting results, leading to the conclusion that light should be regarded as a wave and as a particle. In classical logic, this would result in an antinomy. If something is a wave, it cannot be a particle and \textit{vice versa}. However, from a polycontextural position, this antinomy can be resolved~\citep{guentherDreiwertigeLogikUnd1955}. Both experiments provide a different and measurement-dependent (which here means: dependent on the means of observation or the design of the experiment) view on the object of observation and can hence be considered as a measurement that was done in its own contexture. How can this be formalized in a polycontextural framework? In order to clarify the following analysis, we will utilise an abstraction of the previous experimental setting. Based on the setup of the double-slit experiment, let us stipulate that $O_1$ uses some measurement device that provides information about the momentum (and hence the wavelength) of light, but does not tell us anything about the spatial position. Likewise, $O_2$ uses a device that provides exact information about the spatial coordinates, but does not measure any momentum or wavelength.
Now -- based on the proemial relation -- let us assume that $O_2$ observes $O_1$ conducting its experiment, as sketched in Fig.~\ref{fig:duality}. For $O_1$ the important result of the experiment is that the analysed light behaves like interfering waves. However, $O_2$ observes that $O_1$'s experiment does not make any statements about the spatial properties of light. Hence, $O_2$ does not comment on $O_1$ conclusions but on the process of obtaining it. Additionally, $O_2$ can perform its own experiment and obtains that light consists of particles that have an exact spatial position. Since $O_2$ also observed $O_1$ performing his experiment, $O_2$ must conclude that light has (if measured) a defined wavelength and at the same time can have (if measured) a defined position. During the course of many measurements, $O_1$ and $O_2$ will converge to a common understanding of this fact.\footnote{Note that our stylised experimental setting assumes that the observers measure either only the position or only the momentum and hence conclude that light is either only particle or only wave. It has, however, been established, both theoretically and empirically (e.g. with low-intensity double-slit experiments), that the relationship is more gradual, a phenomenon summarized as quantitative wave-particle duality~\citep{qureshiQuantitativeWaveparticleDuality2016,jacquesDelayedChoiceTestQuantum2008,englertFringeVisibilityWhichWay1996,jaegerTwoInterferometricComplementarities1995}.} However, the process of convergence is only objectively describable by us as a third and external observer.

In the case of the wave-particle duality the incompatibility of the different contextures can be expressed in the following way: In quantum mechanics each observable $A$ (like position or momentum) is associated with an operator $\hat{A}$ that acts on the wave function, and the values of the observable are the eigenvalues of the respective operator. The German physicist Werner Heisenberg discovered that for any measurement it holds~\citep{wernerUncertaintyHeisenbergToday2019}:
\begin{equation}
    \Delta x \Delta p \geq \frac{h}{4 \pi}\,,
\end{equation}
where $\Delta x$ and $\Delta p$ are the standard deviations of the expectation values of the position and momentum operators, respectively (acting on the same wavefunction). Formally, this uncertainty relation (or indeterminacy relation) follows from the commutator relation of conjugated operators. More accessible is, however, the wave function based interpretation of the uncertainty principle. Let us start from the relationship between the (time independent) wave function $\psi (x)$ in position space and the corresponding wave function $\phi (p)$ in momentum space:
\begin{equation}
    \psi(x) = \frac{1}{\sqrt{2 \pi \hbar}} \int_{-\infty}^{\infty}\phi(p) \mathrm{e}^{\frac{i p x}{\hbar}}\mathrm{d}p\,.
\end{equation}
This equation tells us that the wave function in position space is given by adding up states with a defined momentum, weighted by $\phi(p)$. From the above equation follows:
\begin{equation}
    \frac{i}{\hbar} \frac{\mathrm{d}}{\mathrm{d} x}\psi(x) = \frac{1}{\sqrt{2 \pi \hbar}} \int_{-\infty}^{\infty} p\,\phi(p) \mathrm{e}^{\frac{i p x}{\hbar}}\mathrm{d}p\,,
\end{equation}
which can be identified as:
\begin{equation}
    \frac{i}{\hbar} \frac{\mathrm{d}}{\mathrm{d} x}\psi(x) = \hat{p} \psi(x)\,,
\end{equation}
where $\hat{p}$ is the momentum operator in position space. The two views (operator-based, wave function-based) are hence related. The measurement of the momentum projects the wave function into an eigenstate of the momentum operator. As a consequence, the momentum in position space is represented as a sum of multiple eigenstates of the momentum operator.

Heisenberg's uncertainty principle states that any measurement that obtains a particle description and hence a precisely defined position $\Delta x \approx 0$ results in an undefined momentum, i.e. an undefined wavelength. Likewise, any measurement that obtains a defined momentum (a defined wavelength) results in an undefined position and hence an unlocated particle. There is no way to measure exact position and momentum at the same time, or in the words of polycontexturality: there is no way to merge different contextures. They are in a heterarchy. More generally: a physical state can be equally defined in momentum or space coordinates. But once one of these coordinates is chosen, the outcome of the measurement is predetermined and can no longer be in keeping with results obtained in the other coordinates.

While we have shown that the particle-wave duality can be interpreted as a polycontextural phenomenon, it is obvious that modern physics is able to work and to exploit the particle-wave duality even though based on a two-valued logic. As in our previous examples, a transformation can serve as an exchange relation between incompatible logical systems.\footnote{Our notion of transformations among contextures in quantum mechanics is similar to -- but distinct from -- the well-established notion of quantum contextuality, where, given three observables $A$, $B$ and $C$, the result for a measurement of $A$ can be different, depending on whether $A$ is measured together with $B$ or with $C$~\citep{abramskySheaftheoreticStructureNonlocality2011,jaegerQuantumContextualityCopenhagen2019,ghirardiContextualityNonlocalityCounterfactual2009,auffevesContextsSystemsModalities2016}} Starting from the concepts in quantum mechanics it is possible to illustrate the duality described above with the concept of a Fourier transform, which provides an isomorphic mapping between the representation of the wave function in space and momentum coordinates. If the position of a particle is well defined (the uncertainty of the expectation value of the position operator is small), then the probability density of its position is a narrow and localised distribution e.g. a Gaussian function:
\begin{equation}
    g(x) = \frac{1}{\sigma \sqrt{2 \pi}} \textrm{e}^{\frac{-x^2}{2 \sigma^2}},
\end{equation}
with a small variance $\sigma^2$.
The probability density of its momentum -- given by the Fourier transform $\mathcal{F}_x\left[g(x)\right](p)$ of the probability density of its position -- then results in:
\begin{equation}
    \mathcal{F}_x\left[g(x)\right](p) =  \textrm{e}^{-\frac{-p^2 \sigma^2}{2 }},
\end{equation}
a wide and unlocalised distribution with a large variance of $1/\sigma^2$. It should be noted that, as the Fourier transform is a classical (non-quantum) method, this argument just serves as an illustration of the reciprocal relationship of localisations in position space and momentum space.  This hints at an important logical difference between quantum theory and the theory of relativity: the Lorenz transformation eventually allows for a deterministic account of the relative moving observers and could hence -- in principle -- be formulated within an Aristotelian framework: polycontextural logic provides only a good means of visualising the paradoxical situation. Contrarily, in quantum physics, different experiments can yield measured properties that can not be embedded into a single $\sigma$-algebra~\citep{kochenReconstructionQuantumMechanics2015}. As a consequence, observations of different observers can generate irresolvable contradictory claims ~\citep{frauchigerQuantumTheoryCannot2018}, and it is thus not possible to assign truth values jointly to their contrary  propositions~\citep{bruknerNoGoTheoremObserverIndependent2018,bruknerQuantumMeasurementProblem2015}. In quantum theory, it hence becomes unavoidable to incorporate logically inconsistent observer positions, as put forward by polycontextural logic.

\section{Conclusion}
\label{sec:conclusion}
The classical two valued logic requires a sharp separation between observer (subject) and objects. However, in physical reality subject and objects are not necessarily independent. The theory of polycontexturality provides a place value system where subject-object interrelationships can be represented with their real properties. By means of two different examples, we have illustrated how polycontextural concepts can be found in common problems of modern physics. Drawing on the setup of the model of fast-moving observers, we used the viewpoint dependent measurement of time to illustrate Gotthard Günther's theory of polycontexturality. As a second example, we showed how the particle-wave duality of light can be described within a polycontextural form. 
Based on these observations, we argued that a polycontextural understanding of the world is a cornerstone for the correct treatment of context-dependent measurements. 
However, in both examples, the overarching question is why present-day science -- that seems to rely on a two-valued global logic -- can deal with the apparent logical incompatibilities. Our analysis suggests that in modern physics key components of the theoretical formalisms, namely transformations, serve the purpose of incorporating situations that are not describable without a polycontextural understanding. 
These transformations enable an observer in one frame to
 become aware of its own relativity and likewise to correct for it.
Thereby -- under the assumption that the world can be described with a polycontextural framework -- the polycontextural reality is projected on one contexture (and thereby on a two-valued logic).

The significance of polycontextural logic is often viewed in its contributions to formal logic, where it offers a new approach to address inconsistencies of logical systems related to Gödel’s incompleteness theorems~\citep{benseGrundlagenUndExistenzbestimmung1960, kaehrEinubungAndereLekture1979, vonfoersterResponsibilitiesCompetence2003}. The distinct logical systems associated with different contextures allow to distribute contradictions and arrive at non-contradictory systems~\cite[p.87]{gunther1980beitrage}\citep{mahlerMorphogrammatikEinfuhrungTheorie1995}.

With our investigation we emphasize that already the existence of intersubjective differences in measurements and observations seem to call for a description in terms of polycontextural logic. This is well known, e.g., in the social sciences~\citep{jansen_kontexturanalyse_nodate, blaschkeOrganizationCommunicationPerspectives2016}. Here we showed that also in physics the development of theoretical descriptions can be elucidated with the help of polycontextural logic (by identifying transformations as connections among contextures). From our perspective, it is interesting to extend this approach further by investigating theory building in other disciplines with the same set of methods. A field, where this is of particular relevance, is artificial intelligence and machine learning, where devices are trained to classify data. This classification often remain in the realm of binary logic~\citep{ben-david_learnability_2019}. One should also note that we restricted ourselves to only one specific interpretation of Günther's polycontextural logic, namely an observational position. This is the interpretation that Günther follows in his work as well as the interpretation that corresponds best with our everyday understanding. Günther based his polycontextural logic on two theories that he called Morphogrammatic and Kenogrammatic~\citep[p.140]{klagenfurt_technologische_2016}~\citep[p.215]{gunther_beitrage_1976}~\citep[p.109]{gunther1980beitrage}. At its core, these theories allow for an interpretation-free mediation of differences.
Resorting to these theories, it would hence be possible to apply Günther’s formalism without drawing on a specific interpretation. This would e.g. leave open whether contextures are \textit{I} and \textit{You}, or different worlds, or something else. On that basis, our investigation could in principle be extended to other interpretations of (quantum) physics (e.g., the many-worlds interpretation).

Being a natural framework for analyzing the solvability of distributed logical systems, a first implementation of polycontextural algorithms could e.g. be provided for self-organized dynamical processes on graphs~\citep{windt_graph_2010,hutt_perspective_2014}. Additionally, a polycontextural formalism can lead to a formal acceptance and a deeper understanding of undecidable statements as they appear in machine learning~\citep{ben-david_learnability_2019}.

We hope that our attempt to summarize, formalize and apply the basic framework of polycontexturality will help initiate further investigations in this direction.

\section*{Acknowledgement}
We thank Beatrix C. Hiesmayr (University of Vienna) and Thorsten Kohl (TU Darmstadt) for helpful comments and valuable discussions.

\section*{Conflict of Interest}
On behalf of all authors, the corresponding author states that there is no conflict of interest.

%
%
\bibliographystyle{spbasic}
\bibliography{references}

\end{document}